\documentclass[conference, 10pt, a4paper]{IEEEtran}
\IEEEoverridecommandlockouts
\usepackage{amsmath,amssymb,amsfonts}
\usepackage{algcompatible}
\usepackage{graphicx}
\usepackage{textcomp}
\usepackage{xcolor}
\usepackage[a4paper, total={184mm,239mm}]{geometry}
\def\BibTeX{{\rm B\kern-.05em{\sc i\kern-.025em b}\kern-.08em
    T\kern-.1667em\lower.7ex\hbox{E}\kern-.125emX}}
\usepackage[utf8]{inputenc}

\usepackage{algpseudocode}
\usepackage{algorithm}
\usepackage{mathtools}
\usepackage[colorlinks]{hyperref}
\usepackage{indentfirst}
\usepackage[noabbrev,capitalize,nameinlink]{cleveref}
\usepackage[style=ieee,sorting=none,backend=bibtex]{biblatex}
\usepackage{subcaption}
\usepackage{multirow}

\newcommand{\bibliofont}{\fontsize{9}{9}\selectfont}
\defbibenvironment{bibliography}
  {\list
     {\printtext[labelnumberwidth]{%
        \printfield{prefixnumber}%
        \printfield{labelnumber}}}
     {\setlength{\labelwidth}{\labelnumberwidth}%
      \setlength{\leftmargin}{\labelwidth}%
      \setlength{\labelsep}{\biblabelsep}%
      \addtolength{\leftmargin}{\labelsep}%
      \setlength{\itemsep}{\bibitemsep}%
      \setlength{\parsep}{\bibparsep}%
      \bibliofont}
     \sloppy}
  {\endlist}
  {\item}

\usepackage{xcolor}
\usepackage{url}

\usepackage{tikz}

\usepackage{filecontents}
\usepackage{listings}

\definecolor{mGreen}{rgb}{0,0.6,0}
\definecolor{mGray}{rgb}{0.5,0.5,0.5}
\definecolor{mPurple}{rgb}{0.58,0,0.82}
\definecolor{backgroundColour}{rgb}{0.95,0.95,0.92}

\lstdefinestyle{CStyle}{
    backgroundcolor=\color{backgroundColour},
    commentstyle=\color{mGreen},
    keywordstyle=\color{magenta},
    numberstyle=\tiny\color{mGray},
    stringstyle=\color{mPurple},
    basicstyle=\footnotesize,
    breakatwhitespace=false,
    breaklines=true,
    captionpos=b,
    keepspaces=true, 
    numbers=left,
    numbersep=5pt,
    showspaces=false,
    showstringspaces=false,
    showtabs=false,
    tabsize=2,
    language=C
}

\title{SGPRS: Seamless GPU Partitioning Real-Time Scheduler for Periodic Deep Learning Workloads\vspace{-5pt}}

\author{\IEEEauthorblockN{Amir Fakhim Babaei} 
\IEEEauthorblockA{\textit{Department of Electrical \& Computer Engineering,} \\
        \textit{Virginia Tech}\\
        Arlington, Virginia, USA \\
        babaei@vt.edu}\vspace{-25pt}
    \and
    \IEEEauthorblockN{Thidapat Chantem} 
    \IEEEauthorblockA{\textit{Department of Electrical \& Computer Engineering,} \\
        \textit{Virginia Tech}\\
        Arlington, Virginia, USA \\
        tchantem@vt.edu}\vspace{-25pt}
}

\begin{document}
\maketitle

    \begin{abstract}
        Deep Neural Networks (DNNs) are useful in many applications, including transportation, healthcare, and speech recognition. Despite various efforts to improve accuracy, few works have studied DNN in the context of real-time requirements. Coarse resource allocation and sequential execution in existing frameworks result in underutilization. In this work, we conduct GPU speedup gain analysis and propose SGPRS, the first real-time GPU scheduler considering zero configuration partition switch. The proposed scheduler not only meets more deadlines for parallel tasks but also sustains overall performance beyond the pivot point.
    \end{abstract}

    \vspace{-10pt}
    \section{Introduction}\vspace{-0pt}
    The limited processing capability of CPUs is the primary challenge in implementing Deep Learning (DL) techniques, and NVIDIA GPUs are the most popular DNN accelerator. Due to underutilization, the co-location of DNN tasks on GPUs has become inevitable. GPUs can be partitioned spatially or temporally, making spatio-temporal partitioning a state-of-the-art technique to leverage both performance and timeliness \cite{juracy2023cnn, cui2021enable}. 
    Given the limitations of current works to handle periodic real-time DL tasks, we propose \textbf{SGPRS} (\textit{Seamless GPU Partitioning Real-time Scheduler}). 

    \section{\label{sec:background}Task Model}
    We define a task set $S=\{\tau_1, \tau_2, ..., \tau_{|S|}\}$. Each task $\tau_i$ represents a DNN with a DAG structure. The nodes within $\tau_i$ represent the stages (sub-tasks) of the DNN, denoted as $\tau_i^j$. Assume $CP={cp_1, ..., cp_{n_p}}$ represents a context pool with $n_p$ options (CUDA contexts) and $sm$ SMs for each. Let $C_i$ and $C_i^j$ represent the Worst-Case Execution Times (WCETs), while $D_i$ and $D_i^j$ represent the relative deadlines of the task and sub-task, respectively. $D_i$ should be determined initially, while $D_i^j$ is a virtual deadline explained in \cref{sec:wcet}.

    \section{\label{sec:analysis}GPU Speedup Analysis}
\begin{figure}
        \begin{center}
            \includegraphics[width=0.9 \linewidth]{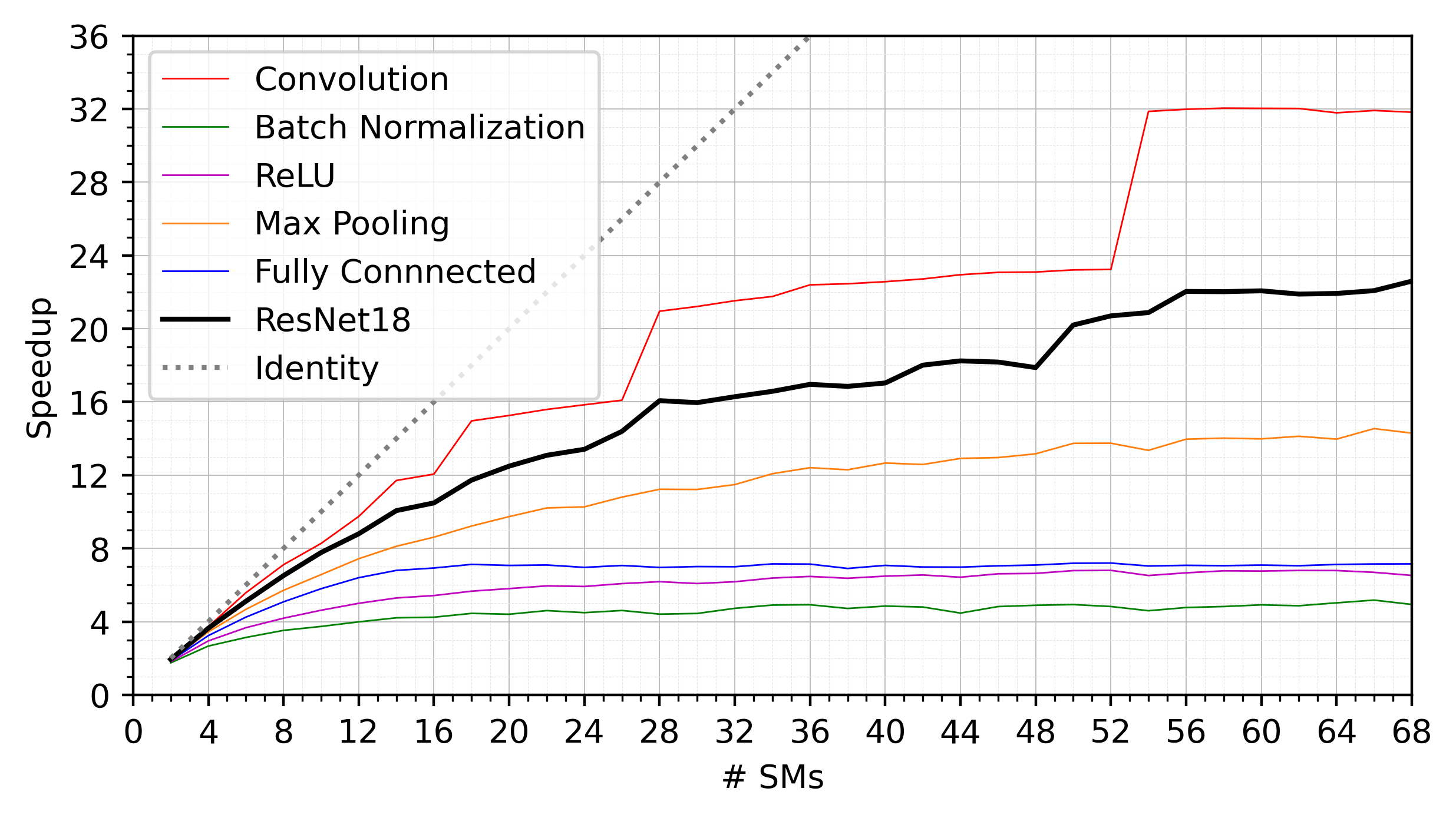}
        \end{center}
        \caption{\label{fig:speedup}Speedup gain for different operations when running in isolation}
    \end{figure}

    A linear speedup is not realistic in GPUs due to their architecture \cite{jiang2019achieving}. To showcase this, we use NVIDIA RTX 2080 Ti which has 68 SMs. ResNet18 \cite{he2016deep} is used as the benchmark DNN to measure the speedup gain of common operations as a function of the number of SMs. The results are illustrated in \cref{fig:speedup}. The \textit{convolution} operation reaches the best speedup gain (32x) followed by \textit{max pooling} (14x). Other operations failed to exceed 7x. The convolution layer dominates the overall speedup behavior of ResNet18 (only 23x) due to the presence of other layers.
    
    \section{\label{sec:scheduler}Real-Time Scheduler Design}
\begin{figure}
        \begin{center}
           \includegraphics[width=0.9 \linewidth]{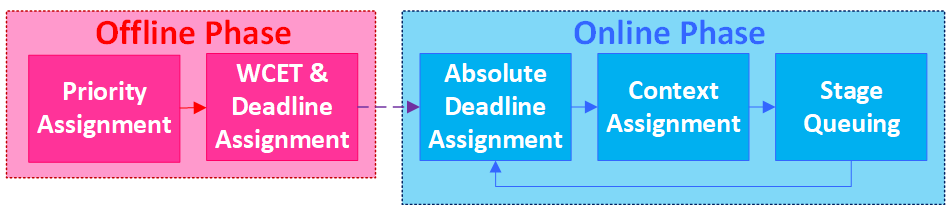}
        \end{center}
        \caption{\label{fig:SGPRS}SGPRS Overview}
    \end{figure}

    Our model philosophy is to consider multi-tenant DNN inference. Popular DNN frameworks such as PyTorch \cite{Paszke_PyTorch_An_Imperative_2019}, can not efficiently benefit from GPU concurrency mechanisms. Our approach results in enhanced efficiency and performance in scenarios with real-time constraints. We use \textit{LibTorch} (C++ PyTorch) in this work. In SGPRS, we propose dividing a network (task) into multiple stages (sub-tasks) to improve flexibility. SGPRS is divided into two main phases, \textit{offline} and \textit{online} as illustrated in \cref{fig:SGPRS}.

    \subsection{\label{sec:offline}Offline Phase}
    \subsubsection{\label{sec:priority}Two Level Priority Assignment}
    We use a two-level offline priority assignment mechanism where the last stage of each task will have a high priority, while the rest of the stages will have a low priority. This helps to meet more deadlines.
    
    \subsubsection{\label{sec:wcet}WCET and Virtual Deadline Assignment}
    The WCETs of each task ($C_i$) and its stages ($C_i^j$) are measured offline. The relative deadline of the stages ($D_i^j$) will be assigned as a portion of the relative deadline of the entire task ($D_i$), proportional to their relative WCET.

    \subsection{\label{sec:online}Online Phase}    
    \subsubsection{\label{sec:arrival}Absolute Deadline Assignment}
    At each instance, the absolute deadline of each stage ($d_i^j$) will be assigned based on their relative virtual deadline ($D_i^j$).

    \subsubsection{\label{sec:contex}Context Assignment}
    Released stages will be assigned to a context based on the following criteria: empty queues first, then the context meeting the deadline with the shortest queue, and if none, the one with the earliest finish time.

    \subsubsection{\label{sec:queue}Stage Queuing}
    We use two high and two low-priority CUDA streams in each CUDA context, meaning a maximum of four stages are allowed in parallel in each context. We also add a third priority level (medium), which is assigned to low-priority stages whose preceding stage has missed its deadline. Stages inside each priority level will be scheduled in the \textit{Earliest Deadline First} order.

    \section{Simulation Results}
    We use the ResNet18 network with a 224x224 input and 30 fps as our benchmark task. We compare the proposed SGPRS alongside a naive approach, which is a simple spatial partitioning scheduler that lacks the context switch and temporal partitioning features. The naive scheduler is the best candidate for comparison because it highlights the benefits of SGPRS over pure spatial partitioning. To reach higher utilization, we use an over-subscribed context pool, meaning, the sum of SMs of all contexts might be larger than the total number of SMs available. Three instances of the proposed scheduler with three over-subscription options are considered. We use \textit{SGPRS\_os} notation where \textit{os} represents the over-subscription level. Also, two metrics have been used for comparison, total FPS and Deadline Miss Rate (DMR). Finally, two scenarios of \textit{two} and \textit{three} context pool options have been chosen.

\begin{figure}
        \begin{center}
            \begin{subfigure}{0.9\linewidth}
                \includegraphics[width=0.9 \linewidth]{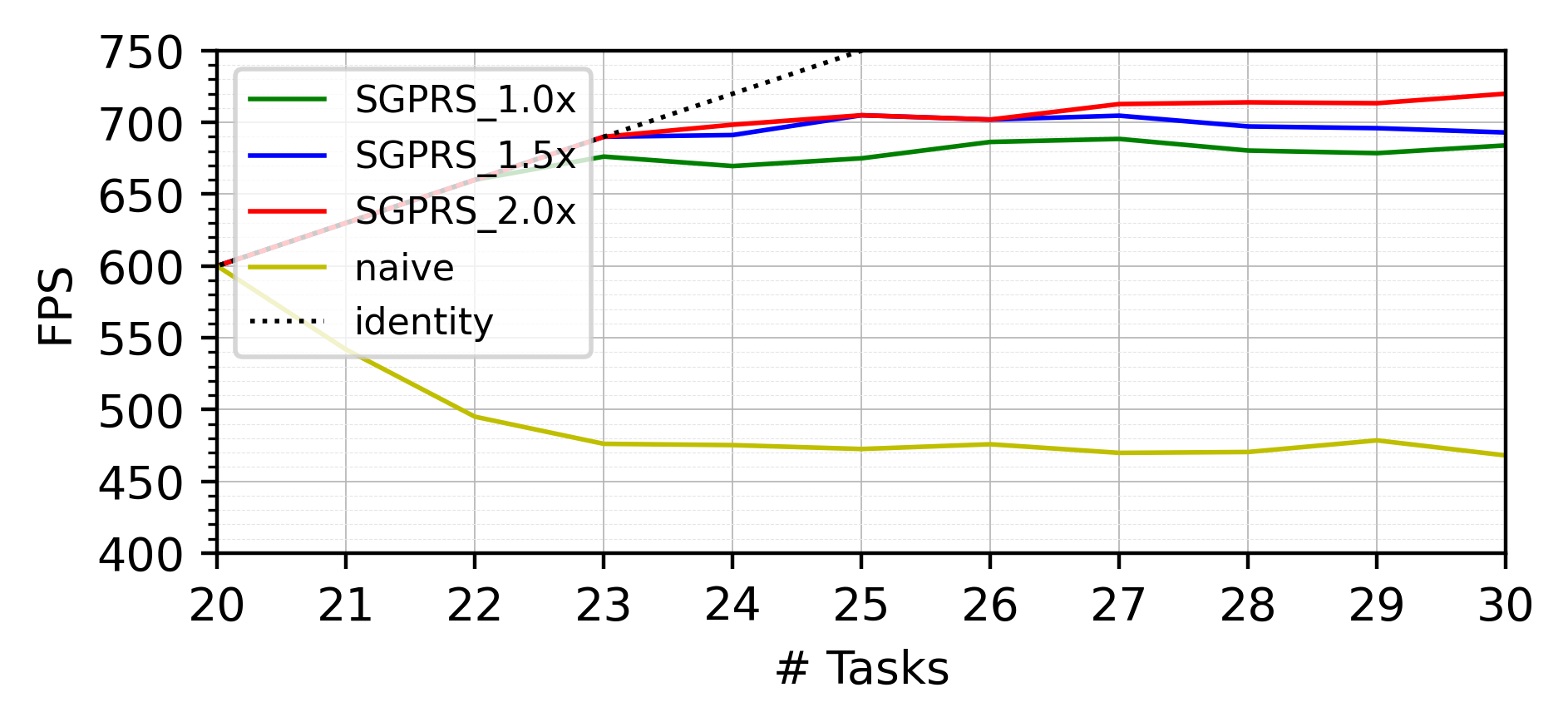}
                \vspace{-7.5pt}
                \caption{\label{fig:fps2}Total FPS reached}
                \vspace{-7.5pt}
            \end{subfigure}
        \end{center}
        \begin{center}
            \begin{subfigure}{0.9\linewidth}
                \includegraphics[width=0.9 \linewidth]{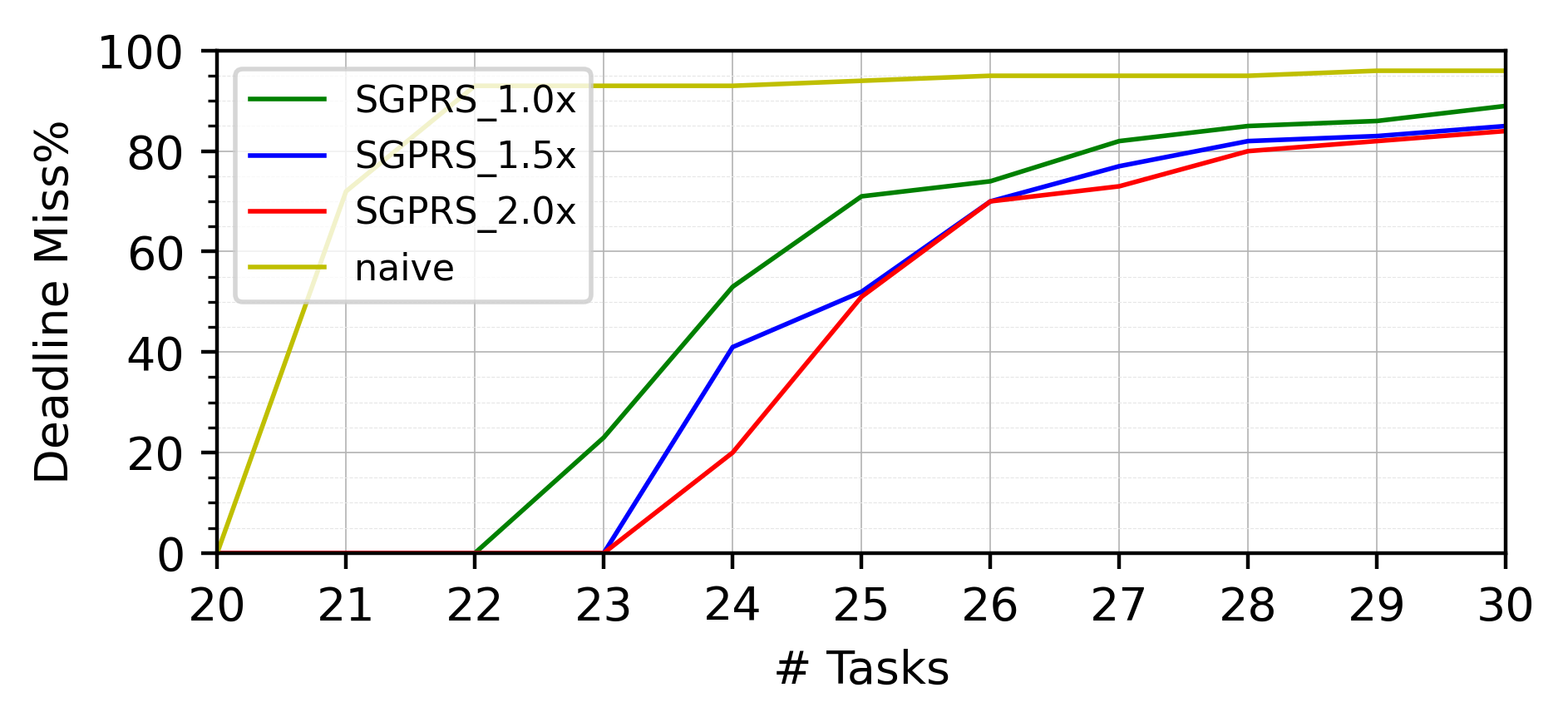}
                \vspace{-7.5pt}
                \caption{\label{fig:miss2}Deadline miss rate}
                \vspace{-7.5pt}
            \end{subfigure}
        \end{center}
        \caption{\label{fig:final2}Result when using 2 contexts (Scenario 1)}
        \vspace{-10pt}
    \end{figure}

\begin{figure}
        \begin{center}
            \begin{subfigure}{0.9\linewidth}
                \includegraphics[width=0.9 \linewidth]{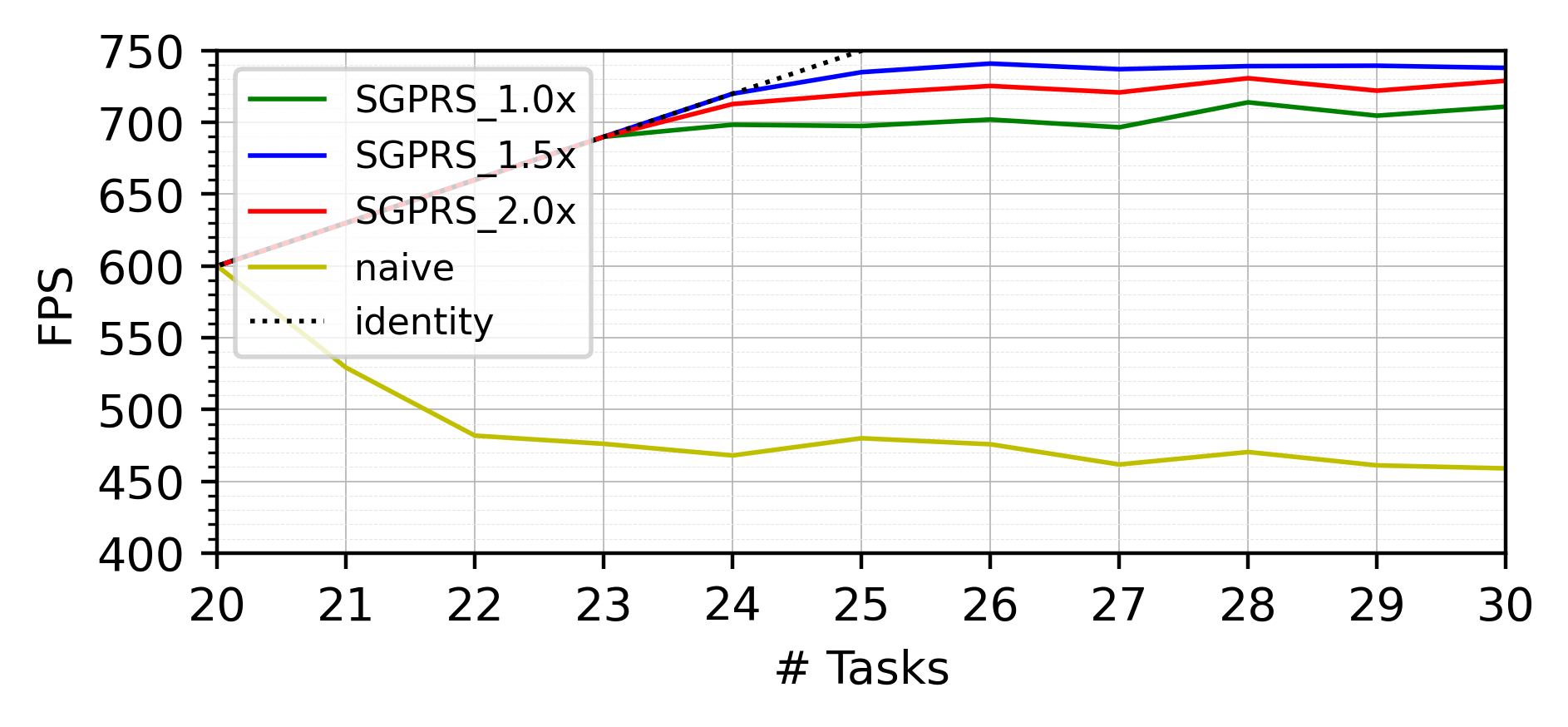}
                \vspace{-7.5pt}
                \caption{\label{fig:fps3}Total FPS reached}
                \vspace{-7.5pt}
            \end{subfigure}
        \end{center}
        \begin{center}
            \begin{subfigure}{0.9\linewidth}
                \includegraphics[width=0.9 \linewidth]{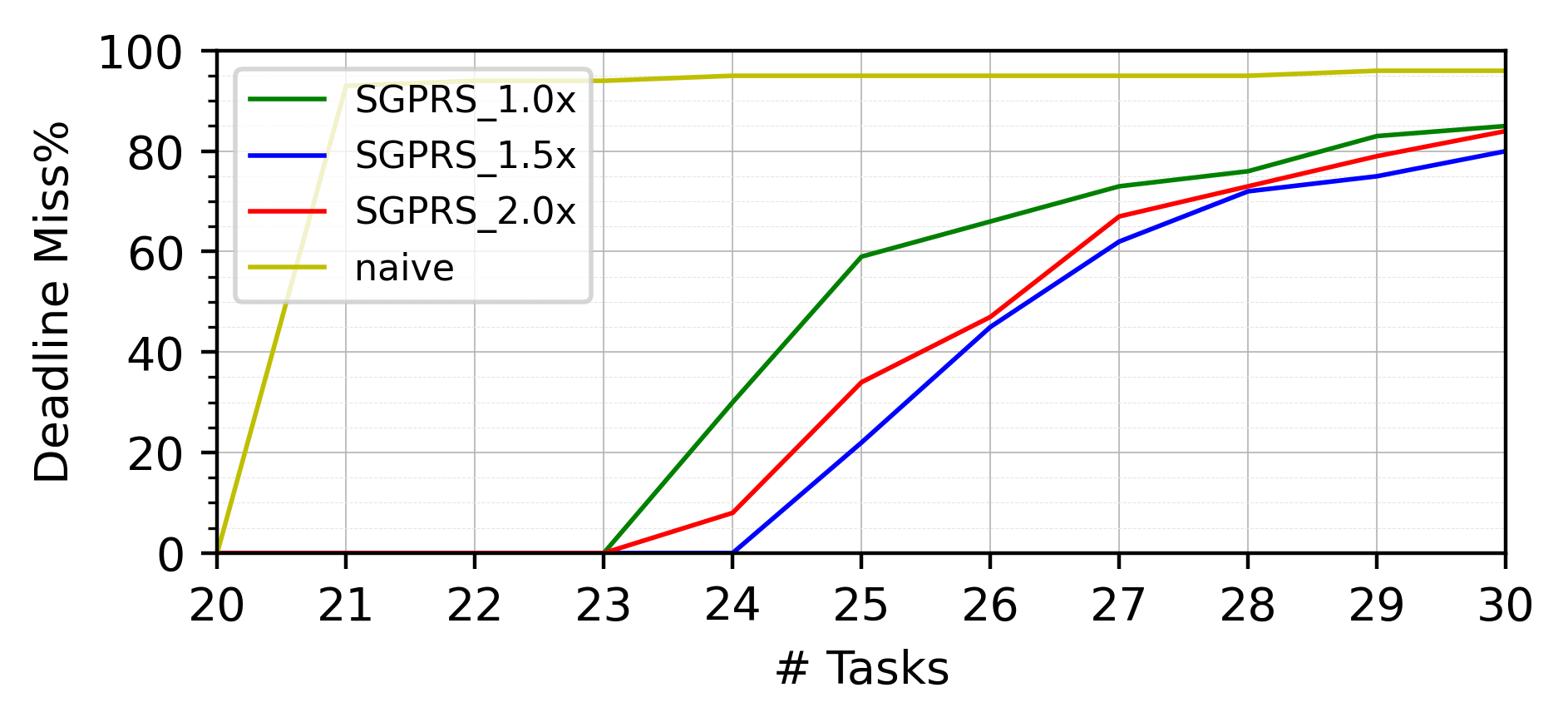}
                \vspace{-7.5pt}
                \caption{\label{fig:miss3}Deadline miss rate}
                \vspace{-7.5pt}
            \end{subfigure}
        \end{center}
        \caption{\label{fig:final3}Result when using 3 contexts (Scenario 2)}
        \vspace{-10pt}
    \end{figure}

    Identical periodic tasks (30 fps) with explicit deadlines with each divided into six stages are considered. We refer to the largest number of tasks that the scheduler can handle without deadline misses as the \textit{pivot point}. In the naive scheduler, for both scenarios (\cref{fig:final2} and \cref{fig:final3}), the pivot point happens much earlier than SGPRS. Also, both total FPS and DMR are degraded drastically after the pivot point. This happens due to the lack of temporal partitioning to prevent the domino effect of deadline misses after the pivot point. SGPRS variations, not only can sustain total FPS, but their DMR increases with a moderate slope. This shows that the lack of proper temporal partitioning leads to excessive contention. The total FPS of the naive scheduler drops to 468 fps and 459 fps for Scenario 1 and Scenario 2. This means 38\% and 36\% FPS drop compared to best-case SGPRS variations.
    
    Even though both scenarios show a similar FPS and DMR pattern, there are some differences. Scenario 2 performs better overall. The best-case pivot point happens at 23 and 24 tasks in Scenario 1 and 2 respectively. Furthermore, in \cref{fig:fps2} the FPS always increases relative to the over-subscription factor, but as represented in \cref{fig:fps3}, the highest over-subscription will not lead to the best performance. Higher over-subscription leads to poor predictability and increased resource contention, which is why in Scenario 2 the 1.5x variation reaches higher (741 fps) compared to 2.0x (731 fps). In Scenario 1, this does not happen because there are too few contexts to leverage the whole GPU and excessive contention is not a problem.

    \printbibliography
\end{document}